\documentstyle[12pt,a4,epsf]{article}
\makeatletter
\def\@cite#1#2{$^{\hbox{\scriptsize{#1\if@tempswa ,#2\fi} }}$}
\renewcommand\section{\@startsection {section}{1}{\z@}%
                                   {-3.5ex \@plus -1ex \@minus -.2ex}%
                                   {2.3ex \@plus.2ex}%
                                   {\normalfont\large\bfseries}}
\renewcommand\subsection{\@startsection{subsection}{2}{\z@}%
                                     {-3.25ex\@plus -1ex \@minus -.2ex}%
                                     {1.5ex \@plus .2ex}%
                                     {\normalfont\normalsize\bfseries}}
\makeatother

\begin{document}
\vspace*{0.5cm}

\begin{center}
{\Large {\bf SIMULATION OF THE IMPACT OF TWO-DIMENSIONAL ELASTIC DISKS}}
\end{center}
\begin{center}
Hisao Hayakawa and Hiroto Kuninaka
\end{center}
\begin{center}
{\it Graduate School of Human and Environmental Studies,
  Kyoto University, Sakyo-ku, Kyoto, 606-8501, Japan}
\end{center}
\begin{quotation}
{\small The impact of a two-dimensional elastic disk with a wall is 
numerically studied. It is clarified that the coefficient of restitution (COR)
decreases with the impact velocity. The result is not consistent with 
the recent quasi-static theory of inelastic collisions even for very slow
impact. The abrupt drop of COR has been found
 due to the plastic deformation of the disk, 
which is assisted by the initial internal motion.
(to be published in Proceedings of 9 th Nisshin Engineering Particle 
Technology International Symposium on 'Solids Flow Mechanics and Their 
Applications' held at Kyoto, 7th-9th January, 2001 )} 
\end{quotation}


\section{Introduction}

\par\indent
The collision of particles with the internal degrees of freedom are
inelastic in general. 
The inelastic collisions are abundant in nature\cite{goldsmith}.
 Examples can be seen
in collisions of atoms, molecules, elastic materials, balls
in sports, and so on.
The study of inelastic collisions 
will be able to be widely accepted as one of fundamental subjects in physics,
because they are almost always discussed 
 in textbooks of  elementary classical mechanics\cite{high}.

 Recent extensive interest in granular
 materials\cite{granules}
 makes 
physicists to recognize fundamental roles
of inelastic collisions. 
In fact, granules consists of macroscopic dissipative particles. 
Therefore, the decision of interaction among particles is obviously
 important. We believe that static interactions among granular
 particles can be described by the theory of
 elasticity\cite{love,landau,johnson,hills}. For example,
 the normal compression may be described by the Hertzian contact
 force\cite{hertz} and the shear force may be represented by the
 Mindline force\cite{mindlin}. The dynamical part related to the
 dissipation, however, cannot be described by any reliable physical theory.  
Thus, the distinct element method\cite{dem} which is one of the
 most popular models to simulate collections of granular particles
 contains some dynamical undetermined parameters.
  In other words, to determine such the parameters is
 important  for both granular physics and fundamental
 physics.

The normal impact of macroscopic materials is 
characterized by the coefficient of restitution (COR)
defined by
\begin{equation}
e=-v_{r}/v_{i} ,
\end{equation}
where $v_i$ and $v_r$ are the relative velocities of incoming and outgoing
 particles respectively.
COR $e$ had been believed to be
  a material constant, since the classical experiment by Newton\cite{newton}.
 In general, however, experiments show that COR
 for three dimensional materials 
is not a constant even in approximate sense
 but
 depends strongly on
 the impact velocity\cite{goldsmith,sonder,1d}.

The origin of the dissipation in inelastic collisions is the transfer of
the kinetic energy of the center of mass 
into the internal degrees of freedom during the impacts.
Systematic theoretical investigations of the impact have begun with the paper
by Kuwabara and Kono\cite{kuwabara}. 
Taking into account the
viscous motion among the internal degrees of freedom, they derived 
the equation of the macroscopic deformation.
Later, Brilliantov {\it et al.}\cite{brilliantov96} and
Morgado and Oppenheim\cite{morgado}  derived the identical equation
to eq.(\ref{oppenheim}). In
particular, the derivation by Morgado and Oppenheim\cite{morgado} 
is based on the standard technique of nonequilibrium statistical
mechanics to extract the slow mode among the fast many modes which can
be regarded as the thermal reservoir.
Furthermore, Brilliantov {\it et al.}
compared their theoretical results with experimental results\cite{brilliantov}.
Thus, the quasi-static theory has been accepted as reasonable one.

On the other hand, Gerl and Zippelius\cite{gerl} performed the
 microscopic simulation of the two-dimensional collision of an elastic disk
 with a wall.  Their simulation is mainly based on the mode expansion of 
an elastic disk under the force free boundary condition. Then, they
 solve Hamilton's equation determined by the elastic field and the
 repulsive potential to represent the collision of two disks. 
Their results show that COR decreases with the impact velocity, which
strongly depends on Poisson's ratio. For high velocity of the impact 
they demonstrate the macroscopic deformation has left after the
collision is over. 
Although it is not easy to discuss
the impact with the very low impact velocity from their method,
 their analysis may suggest the possibility of a complicated relation
 between 
 the quasi-static theory of impact\cite{kuwabara,brilliantov96,morgado} 
and their microscopic
 simulation\cite{gerl}.
Thus, we have to clarify the relation between two typical approaches.

In this paper, we will perform the microscopic simulation of the impact 
of a two dimensional elastic disk with a wall. We introduce two methods
of simulation; One is based on the lattice model (model A)
and another is continuum model (model B) which is identical to that by Gerl and
Zippelius\cite{gerl}. Through our simulation, we will demonstrate that 
(i) the effect of temperature (the initial internal motion) is important,
(ii) COR is suddenly dropped by the plastic
deformation which is enhanced by the initial temperature, and (iii) the
continuum model (model B) does not recover the results predicted by 
the quasi-static
theories in the low impact velocity\cite{kuwabara,brilliantov96,morgado}.      

The organization of this paper is as follows. In the next section, we
will briefly review the outline of 
quasi-static theory\cite{kuwabara,brilliantov96,morgado}. 
In section 3, we will explain model A and model B which is equivalent to 
the model by Gerl and Zippelius\cite{gerl} of our simulation.
In section 4, we will show the result of our simulation and discuss the
validity of quasi-static theory. In section 5, we discuss our results,
in particular, about the plastic deformation by the impact and its
origin. In section 6, we will summarize our result.

\section{Quasi-static theory}

In this section, we briefly explain the outline of quasi-static theory.
One purpose of this section is to summarize the two-dimensional version
of quasi-static theory which may not be mentioned in any articles
explicitly. 

At first, let us summarize the three dimensional result, in which
the equation of the macroscopic deformation is given by
\begin{equation}\label{oppenheim}
\ddot h=-k h^{3/2}-\gamma \sqrt{h}\dot h
\end{equation}
in a collision of two spheres, 
where the macroscopic deformation $h$ is given by
 $h=R_1+R_2-|{\bf r}_1-{\bf r}_2|$ with the radius $R_i(i=1,2)$ and the position
of the center of the mass ${\bf r}_i$ of $i$ th
particle. $\dot h$ and $\ddot h$ are respectively $dh/dt$ and $d^2h/dt^2$.
 $k$ and $\gamma$ are unimportant constants. The first term of the right
hand side in eq.(\ref{oppenheim})
 represents the Hertzian contact force\cite{love,landau,johnson,hertz}
 and the second term is
the dissipation due to the internal motion.  

The simplest derivation of eq.(\ref{oppenheim})
 is that by Brilliantov {\it et
al.}\cite{brilliantov96}, though we
also check it validity by the alternative methods. Taking into account
the limitation of the length of this paper, we follow the argument by
them.  

The static stress tensor in the two-dimensional 
linear elastic material can be represented by
\begin{equation}\label{ela}
{\sigma^{(el)}}_{ij}=2\mu (\epsilon_{ij}-\delta_{ij}\epsilon_{ll}/2)+K\delta_{ij}\epsilon_{ll}
\end{equation}
where $\mu$ and $K$ are respectively the shear modulus 
and the compressional modulus, and $\epsilon_{ij}$ is given by 
\begin{equation}
\epsilon_{ij}=\frac{1}{2}\left(\frac{\partial u_i}{\partial x_j}+
\frac{\partial u_j}{\partial x_i}\right)
\end{equation}
with the displacement field $u_i$. 

 The two dimensional Hertzian contact law\cite{johnson,gerl} is given by
the relation between the macroscopic deformation of the center of mass
 $h$ and the elastic force $F_{el}$  as
\begin{equation}
h\simeq-\frac{F_{el}}{\pi Y}\{ \ln \left(\frac{4 \pi Y
        R}{F_{el}\left(1-\sigma^{2}\right)}\right)-1\} ,\label{eq:hertz}
\end{equation}
where $Y$, $\sigma$ and $R$ are the Young modulus, Poisson's ratio and
the radius of the disk without deformation, respectively.
Equation (\ref{eq:hertz}) can be derived from the stress tensor
(\ref{ela}) with the standard treatment of linear elastic theory.
Note that $h$ satisfies $h=R-y_0$ with  the position of the
 center of mass $y_0$\cite{gerl}. 

For small dissipation, as in the textbooks\cite{landau}, the dissipative 
stress tensor due to the viscous
motion among internal motions is given by
\begin{equation}\label{vis}
{\sigma^{(vis)}}_{ij}=2\eta_1 (\dot\epsilon_{ij}-\delta_{ij}\dot\epsilon_{ll}/2)+\eta_2\delta_{ij}\dot\epsilon_{ll}
\end{equation}
as in the case of viscous fluid, where $\dot\epsilon_{ij}$ is the time
derivative of $\epsilon_{ij}$, $\eta_{i}$ ($i=1,2$) is the viscous
constant.

Brilliantov {\it el al.}\cite{brilliantov96}
 assumed that the velocity of deformation field is
governed by the macroscopic deformation, {\it i.e.},
$\dot u_i\simeq \dot h(\partial
u_i/\partial h)$. 
Since in the limit of $v_i\to 0$  we may replace eq.(\ref{eq:hertz})
by $F_{el}\simeq -\pi Y h/\ln(4R/h)$\cite{gerl}. 
Thus, with the aid of the assumption by Brilliantov {\it et
al.}\cite{brilliantov96}, (\ref{ela}) and (\ref{vis}), it is easy to derive
the two dimensional version of quasi-static theory  as
\begin{equation}\label{2d-quasi} 
F_{tot}\simeq -\frac{\pi Y  h}{\ln(4R/h)}- A \frac{\pi Y \dot h}{\ln(4R/h)} , 
\end{equation}
where $A$ is an unimportant constant. 
This result can be derived by various other method. Thus, we will compare the result of our simulation with eq.(\ref{2d-quasi}).

\section{Our models}

Let us explain the details of our model. In both models, the 
wall exists at $y=0$, and the center of mass keeps the
position at $x=0$. The disk approaches from $y > 0$ region
and is rebounded by the wall.

\subsection{Model A}

\begin{figure}[htbp]
 \epsfxsize=6.5cm
 \centerline{\epsfbox{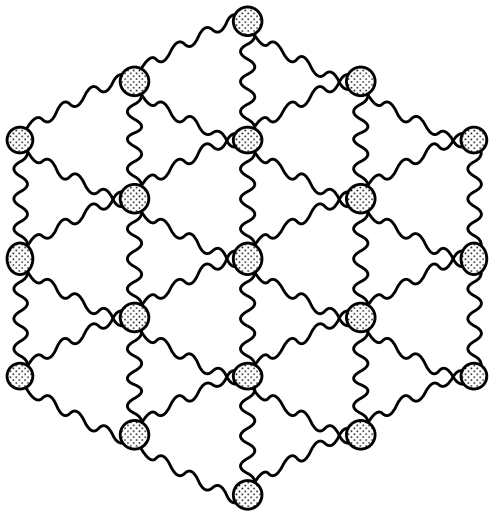}} 
\begin{quotation} \caption{\small A schematic figure of a disk used in model A.
}\end{quotation}
 \label{network}
\end{figure}

The disk in model A
 consists of some mass points (with the mass $m$) on the triangular lattice. All the mass points are combined
 with linear springs with the spring constant $\kappa$.
 In the limit of large number of the mass points, this disk
 corresponds to the continuum circular disk with 
 the Young's modulus 
  $Y=2 \kappa / \sqrt{3}$ and Poisson's ratio $1/3$\cite{hoover}.
The position of each mass point of model A is governed by the
 following equation:
\begin{equation}\label{modelA}
m \frac{d^2 {\bf r}_p}{dt^2}=-\kappa\sum_{i=1}^6(d_0-|{\bf r}_p-{\bf r}_i|)
\frac{{\bf r}_p-{\bf r}_i}{|{\bf r}_p-{\bf r}_i|}
+\widehat{y} a V_{0} e^{-a y_{p}}
\end{equation}
 where $d_0$ is the lattice constant, ${\bf r}_{i}$ is the
 position of the nearest neighbor mass points of
 ${\bf r}_{p}$, $m$ is the mass of the mass points,
 $y_{p}$ is the $y$ coordinate of ${\bf r}_{p}$, and
 $\widehat{y}$ is the unit vector of $y$ direction.
 Note that the directional projection of the linear spring force in model A
can cause the nonlinear deformation. 
The
 wall potential is  
given by
$V_{0}e^{-ay} $
, where $V_{0}=a/2$ and $a=100 /d_0$  for
 model A.
The number of the mass points is fixed to 1459 in model A, since the rough
 evaluation of convergence of the results has been checked in this model.

\subsection{Model B}

In this subsection, we introduce model B which is originally proposed by 
 Gerl and Zippelius\cite{gerl}. Although the details of this model can
 be checked in their paper, we present the minimum description of this
 model to understand the setup of our simulation.

 Gerl and Zippelius\cite{gerl}
 analyze Hamilton's equation ;
\begin{equation}\label{H's equation}
\dot P_{n,l}=-\frac{\partial H}{\partial Q_{n,l}} ;\quad
\dot Q_{n,l}=\frac{\partial H}{\partial P_{n,l}}
\end{equation}
under the Hamiltonian
\begin{equation}\label{hamiltonian}
H=\frac{p_0^2}{2M}+\sum_{n,l}^N(\frac{P_{n,l}^2}{2M}+{1}{2}M\omega_{n,l}^2
Q_{n,l}^2)+V_0\int_{-\pi/2}^{\pi/2}d\phi e^{-ay(\phi,t)} .
\end{equation}
Here $Q_{n,l}$ is the expansion coefficient of the 2D elastic deformation 
field  in the polar coordinate ${\bf u}=(u_r,u_{\phi})$ 
\begin{equation}
(u_r(r,\phi),u_{\phi}(r,\phi))=\sum_{n,l}
Q_{n,l}(u_r^{n,l}(r)\cos n\phi,u_{\phi}^{n,l}(r)\sin n\phi),
\end{equation}
where $u_r^{n,l}(r)R=A_{n,l}\displaystyle\frac{dJ_n(k_{n,l}r)}{dr}+
n B_{n,l}\displaystyle\frac{J_n (k'_{n,l}r)}{r}$
and 
$u_{\phi}^{n,l}(r)R=-n A_{n,l} \displaystyle\frac{J_n(k'_{n,l}r)}{r}-
B_{n,l}\displaystyle\frac{dJ_{n,l}(k_{n,l}r)}{dr}$ 
with the radius of the disk and the Bessel function of the $n-$th order $J_n(x)$.
Here $k'_{n,l}=k_{n,l}\sqrt{2(1+\sigma)/(1-\sigma^2)}$ and $k_{n,l}$ is 
the solution of
\begin{eqnarray}
& &(1-\sigma^2)(1-n^2)\kappa\kappa'^2J_{n-1}(\kappa)J_{n-1}(\kappa')
+ \kappa^2[\kappa^2-2n(n+1)(1-\sigma)]J_n(\kappa)J_n(\kappa')
\nonumber \\
& & + (1-\sigma)[\kappa^2-(1-\sigma)(1-n^2)n][\kappa J_{n-1}(\kappa)J_n(\kappa')
+\kappa'J_{n-1}(\kappa')J_n(\kappa)]=0
\end{eqnarray}
with Poisson's ratio $\sigma$, $\kappa=k_{n,l} R$ and
$\kappa'=k'_{n,l}R$, which is given by the boundary condition. 
Thus, for fixed $n$ there are infinitely many solutions $k_{n,l}$
and $\omega_{n,l}=k_{n,l}\sqrt{Y/\{\rho(1-\sigma^2)\}}$
 numbered by
$l=0,1,\cdots,\infty$.
$A_{n,l}$ and $B_{n,l}$ are
determined by 
\begin{eqnarray}
& & -A_n[\frac{(1-\sigma)}{R}\frac{dJ_n(k_{n,l}R)}{dR}+
(k_{n,l}^2-\frac{(1-\sigma)}{R^2} n^2)J_n(k_{n,l}R)] \nonumber \\
& &+n B_n(1-\sigma)[
\frac{1}{R}\frac{dJ_n(k'_{n,l}R)}{dR}-\frac{J_n(k'_{n,l}R)}{R^2}]=0
\end{eqnarray}
 and
$\int_0^Rdr r \{{u_r^{n,l}}^2+{u_{\phi}^{n,l}}^2\}=R^2$. 
 $P_{n,l}$ is
 the canonical momentum.
$y(\phi,t)$ is the shape of the elastic disk in the polar
coordinate;
\begin{equation}
y(\phi,t)=
y_0(t)+
\sum_{n,l}Q_{n,l}(C_{n,l}\cos(n\phi)\cos\phi-S_{n,l}\sin(n\phi)\sin\phi)
\end{equation}
with the position of the center of mass $y_0(t)$ and constants $C_{n,l}$ 
and $S_{n,l}$ determined by the maximal radial and tangential
displacement at the edge of the disk as $C_{n,l}=u_r^{n,l}(R)$ and 
$S_{n,l}=u_{\phi}^{n,l}(R)$.
 $M$ is the mass of the disk, and the momentum of the
center of the mass $p_0=M\dot{y_{0}}$ satisfies 
$ \dot p_0=- (\partial H/\partial y_0)$
, $V_0$ and $a$ are parameters to express
 the strength of the wall potential.

For the simulation of a pair of identical disks, they
extrapolate the results of their simulation to $a\to \infty$ and $N\to \infty$
with the total number of modes $N$.  
We only adopt $N=1189$ ($n\le 50$ and $\kappa_n\le 50$)or $N=437$
($n\le 30$ and $\kappa_n\le 30$), $V_0=a/2$ and
 $a=500/R$ with the radius of the disk $R$.


\subsection{Parameters in both models}

For the sake of simplicity and comparison between two
different models, 
we only simulate the case of Poisson's ratio $\sigma=1/3$. 
The numerical scheme of the integration of model A is the classical
 fourth order Runge-Kutta method with 
$\Delta t=1.6\times 10^{-3} \sqrt{m/\kappa}$. For model B,
 we adopt the fourth order 
symplectic integral method with $\Delta t=5.0\times
 10^{-3}R/c$ with $c=\sqrt{Y/\rho}$
for model B where $Y$ 
is Young's modulus and $\rho$ is the density. In both models, we have checked the conservation of
 the total energy. 

 We also investigate the impact with the finite 
 temperature. The temperature is introduced as follows:
In model A, we prepare the Maxwellian for the initial velocity
 distribution of mass points, 
where the positions of all mass points are located  at their equilibrium
 positions.
 From the variance of the Maxwellian we can 
 introduce the temperature as a parameter.
 To perform the simulation,
we prepare 10 independent samples obeying the Maxwellian with the
 aid of normal random number.
In model B, we prepare samples in which the absolute value of each mode
 satisfies equipartition law exactly. The sign of each mode is
 assumed to be at
 random with the aid of the uniform random number. 
>From the equipartition law we can introduce the temperature as
 a parameter of simulation, too.

The summary of differences between model A and B is as follows:
(i) All of the mass points in model A interact with  the wall
 but,  in model B, 
only exterior boundary has the influence of the potential as in
 (\ref{hamiltonian}).
(ii) Model A can have nonlinear deformations,
 but model B is based on the theory
 of linear elasticity. 
(iii) Model A can express some plastic deformations, but model B cannot.
(iv) Model A has the six folds symmetry but model B has the rotational
 symmetry.
   
\section{Results}


Now, let us explain the details of the result of our simulation.
In the first subsection, we will introduce the result at $T=0$ and in
the second subsection, we will show the result at finite $T$.

\subsection{Simulation at $T=0$}
 
\begin{figure}
[htbp]
 \epsfxsize=12cm
 \centerline{\epsfbox{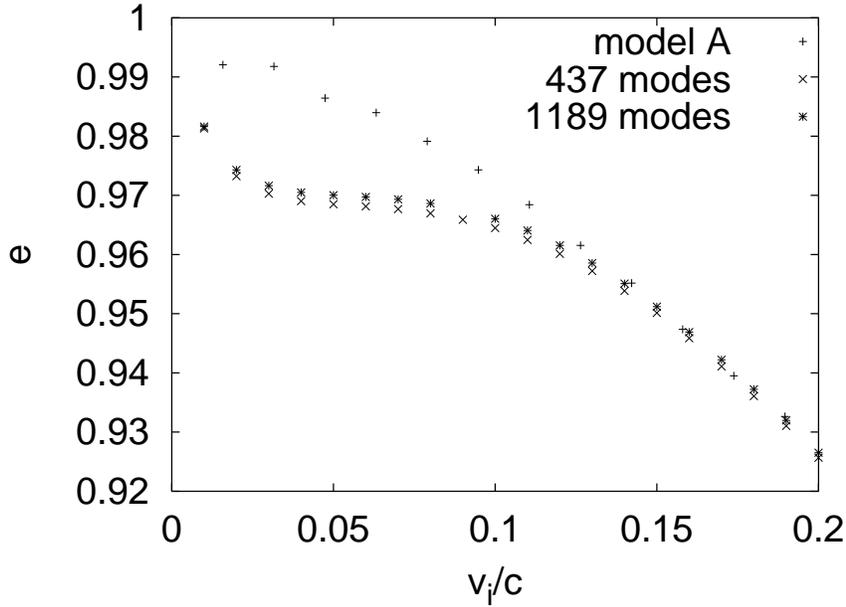}} 
\begin{quotation}
 \caption{\small Coefficient of restitution for normal collision of the 
Model A and Model B as a function of impact velocity, where
 $c=\sqrt{Y/\rho}$ with the Young's modulus $Y$ and the density $\rho$.}
\end{quotation}
 \label{rest}
\end{figure}

At first, we carry out the simulation of model A and model B 
with the initial condition at $T=0$ ({\it i.e.} no internal
motion).
Figure \ref{rest} is the plot of 
the COR against
the impact velocity for both model A and model B. For
  model B, we show the results of $437$ modes and $1189$ modes which
  clearly demonstrates the convergence of the result for the number of modes.
 When impact velocity $v_{i}$ is larger than $0.1c$ with $c=\sqrt{Y/\rho}$,
 the value of COR of
 model A is almost identical to that of model B.
 Each line decreases smoothly as impact velocity increases.
At present, we do not know the reason why the significant difference
 between two models exists at low impact velocity.

Second,
 we investigate the  force acting on the center of mass of the disk
caused
 by the interaction with the wall in model B.
In the limit of $v_i\to 0$ we expect that the Hertzian contact 
theory can be used\cite{landau,johnson,gerl}.
The small amount of transfer from the translational motion
 to the internal motion is  the macroscopic dissipation.
Thus, we  can check the validity of quasi-static 
approaches\cite{kuwabara,brilliantov96,morgado}
 from our simulation by the difference between the observed force acting on
 the center of mass and the Hertzian contact force.

\begin{figure}[htbp]
 \epsfxsize=15cm
 \centerline{\epsfbox{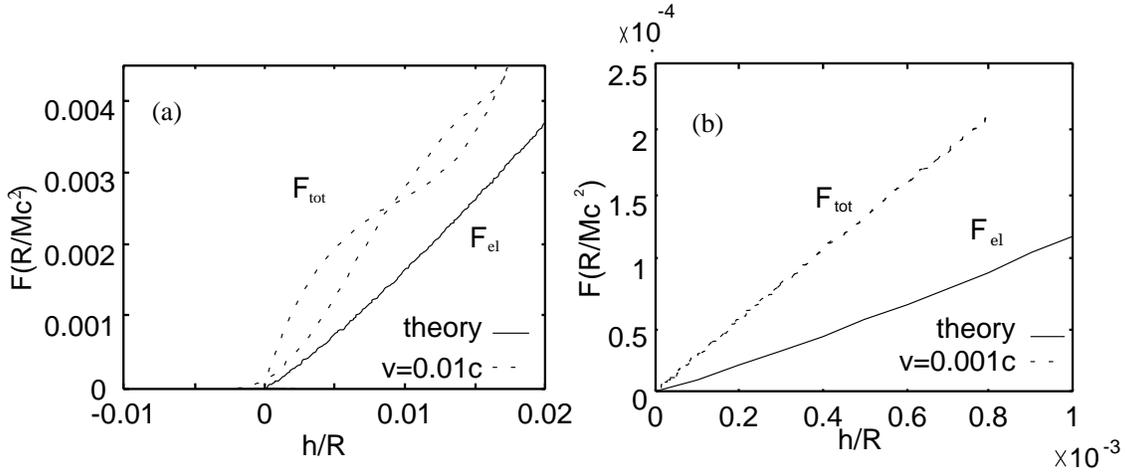}}
\begin{quotation}
 \caption{\small The comparison of the Hertzian force in
   eq.(\ref{eq:hertz}) with our simulation at $v_{i}=0.01c$
  (a) and  $v_i=0.001c $(b) at $T=0$ in model B.}
\end{quotation}
 \label{fel2}
\end{figure}
 
If $h$ is given, we can calculate the
 elastic force by solving eq.(\ref{eq:hertz}) numerically.
Figure 2 is the comparison with
our simulation in model B (1189 modes) and the Hertzian contact theory 
(\ref{eq:hertz}) which is given by
 the solid lines.
 The result of our simulation at the impact velocity 
$v_i=0.01c$ with $c=\sqrt{Y/\rho}$ shows the beautiful hysteresis as 
suggested in the simulation at $v_i=0.1c$ in ref.\cite{gerl}.
This means the compression and rebound are not symmetric.
The hysteresis curve is still self-similar even at
$v_i=0.04c$ but the loop becomes noisy at $v_i=0.1c$.
\begin{figure}[htbp]
\epsfxsize=10cm
\centerline{\epsfbox{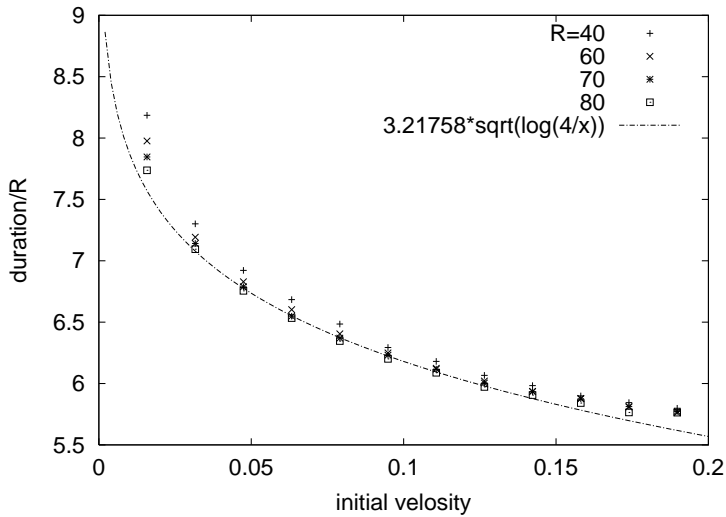}}
\begin{quotation}
\caption{\small The plot of  contact time versus the impact velocity. $R$
represents the radius of the disk, in which $R=40$,$60$,$70$ and $80$
 correspond to the
number of mass points 5815, 13057, 17761 and 23233, respectively.
 The dotted line is fitting curve based on
the quasi-static theory. }
\end{quotation}
\label{contact}
\end{figure}

For very low impact velocity $v_i=0.001c$, the hysteresis loop
disappears but the total force observed in our simulation is almost a
linear function of $h$ which is deviated from the Hertzian contact theory
and quasi-static theory (\ref{2d-quasi}).
In particular, the turning point at $\dot F=0$ is deviated
from the Hertzian 
curve (the solid line). This deviation is clearly contrast to the quasi
static theory, because the dissipative force in the theory
in eqs.(\ref{oppenheim}) and (\ref{2d-quasi}) 
must be zero at the turning 
point at which $\dot h=0$ should satisfy.
This tendency is invariant even for the simulation of model A, though
the data becomes noisy. 
The linearity of the total repulsion force is not
surprising, because
$e^{-a y(\phi,t)}$ in the potential term in
eq.(\ref{hamiltonian}) can be expanded by series of 
$Q_{n,l}$ for very slow impact. 
Although we cannot judge whether the model itself is not appropriate for 
slow impact or the quasi-static theory is wrong, our result clearly means 
that the validity of the quasi-static theory cannot be supported by our
microscopic simulation. However, the validity of the contact time $\tau$ in
the impact evaluated by the quasi-static theory \cite{gerl} can be evaluated as $\tau \simeq (\pi R/c) \sqrt{\ln(4c/v_{i})}$ has been
confirmed by the results of our simulation of model A (Fig.4). Thus, at least, the relation between dynamical
impact theory and quasi-static elastic theory is not trivial at present.

\subsection{Simulation at finite $T$}

Now, let us show the results of our simulation at finite $T$ which has
significant differences from those at $T=0$ in both low and large impact 
velocities. In this sense, we have much 
room to study this process at finite $T$ systematically.

For small impact velocity, COR at finite $T$ becomes larger 
 than that at $T=0$ in both models.
 In some trials COR becomes larger than 1.
It is an interesting result to extract work of this system
 from thermodynamical point of view. The details of the temperature
 effect at the slow impact will be reported elsewhere.

\begin{figure}[htbp]
 \epsfxsize=10cm
 \centerline{\epsfbox{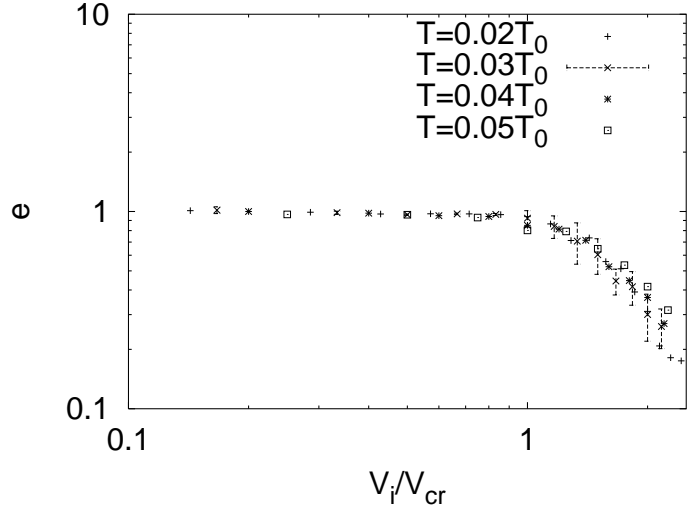}} 
\begin{quotation} \caption{\small 
The relation the coefficient of restitution and
   the impact velocity rescaled by the critical velocity for
   each temperature. Curves are plotted in the log-log
   scale. The temperature is scaled by $T_{0}=mc^2/k_B$ with 
the mass of the mass points $m $ and Boltzmann constant
$k_B$. Note that the error bars are plotted only in the case 
$T/T_{0}=0.03$ but is the same order even at other $T$.}
\end{quotation}
 \label{scale2}
\end{figure}

\begin{figure}[htbp]
 \epsfxsize=10cm
 \centerline{\epsfbox{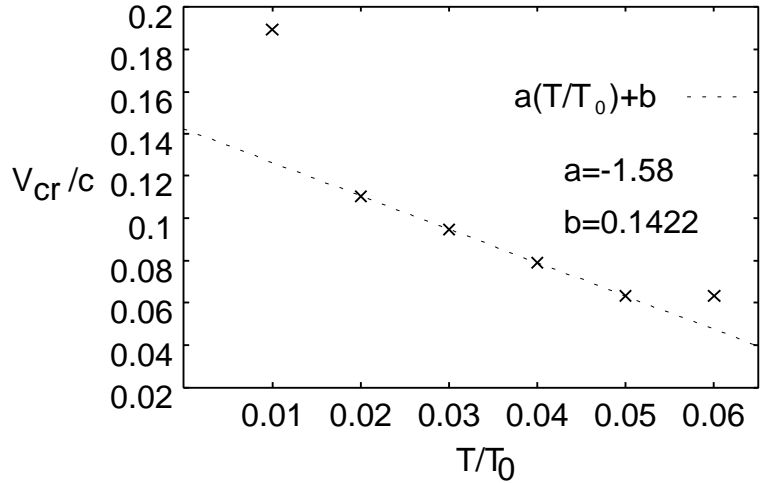}} 
\begin{quotation}
 \caption{\small The plot of the initial temperature and the
   critical velocity causing the plastic
   deformation. $v_{cr}/c=a(T/T_{0})+b$ is the fitting curve
   line from the data between $T/T_{0}=0.02$ and $0.05$.}
\end{quotation}
 \label{vcr}
\end{figure}

For large impact velocity, we do not observe any definite temperature effect in
model B but we find drastic drop of COR in model A.
It seems that COR can be on a universal curve when the impact velocity
is scaled by the critical velocity 
above which COR drops abruptly (Fig.\ref{scale2}). 
 The relation between the critical velocity and the initial 
 temperature at the intermediate impact velocities 
is shown in the Fig.\ref{vcr}.
 The critical velocity seems to obey a linear function of $T$,
though the data is not on the very slow and the very fast impacts. 

\section{discussion}

\begin{figure}[htbp]
 \epsfxsize=6cm
 \centerline{\epsfbox{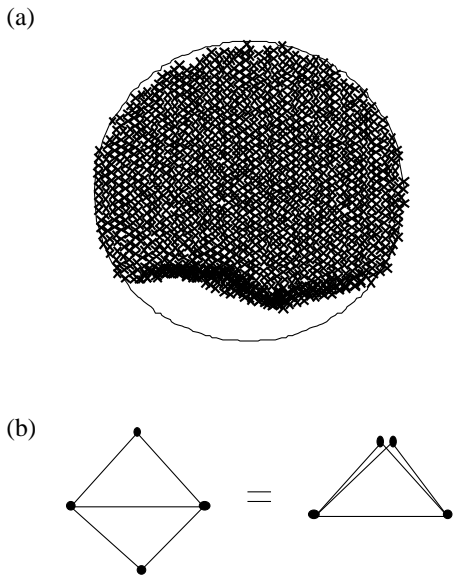}} 
\begin{quotation}
 \caption{\small
(a) Plastic deformation of model A with $v_{i}=0.22$
   at $T=0.03$. The solid circle represents the initial
   circle. The cross points are positions of the mass points 
   after the collision. (b) Two configurations are
 energetically equivalent.}
\end{quotation}
 \label{deform}
\end{figure}

 We investigate what happens in the disk
 above the critical velocity and find the existence of plastic
 deformation of the disk (Fig.\ref{deform}(a)).
Actually, there is no energy differences between two configurations in
 Fig.\ref{deform}(b) 
which can occur after the strong compression during the impact
 but cannot be released  after the impact is over. 
It is well known that the plastic deformation causes the drop of 
COR\cite{johnson}.

Following the description in ref.6, 
let us explain the dimensional analysis of the two-dimensional plastic
deformation. From two-dimensional Hertzian law (\ref{eq:hertz}) we evaluate
     $h \sim a_{0}^2/R$ where $a_{0}=\sqrt{4 F_{el} R/(\pi
       Y^{*})}$ with the elastic force $F_{el}$, the radius
     without contact $R$ and the effective Young's modulus
     $Y^{*}=Y/(1-\sigma^{2})$ with Poisson's ratio $\sigma$
     is the contact length in Hertzian
     law\cite{johnson}. The work for the compression of the
     disk $W$ is $W=(1/2)Mv_{i}^2 \sim
     \int_{0}^{h^{*}}dhF_{el} \sim 
     \int_{0}^{a_{0}^{*}}da_{0}a_{0}/R$, where $M$ and
     $v_{i}$ are the mass of the disk and the impact
     velocity, respectively. $h^{*}$ and $a_{0}^{*}$ are
     respectively the maximal compression and and the
     maximal contact length. Here we neglect the logarithmic 
     correction and unimportant numerical
     factors. Introducing the mean contact pressure during
     dynamical loading $p_{d}$ which satisfies $p_{d} \sim
     F_{el}/a_{0}$, $W$ can be evaluated by $W \sim p_{d}(a_{0}^{*})^{3}/R$. From $W\sim Mv_{i}^{2}$ we can
     express $a_{0}^{*} \sim (Mv_{i}^{2}R/p_{d})^{1/3}$.

Let us assume that the impact exceeds the yield pressure for 
the plastic deformation. In such the case, the deformation
during rebound is frozen. Thus, the work in a rebound is $W' 
\sim F_{*}h^{*}$ where $F_{*}$ is the maximal force during
the impact. From $h^{*}\sim F_{*}/Y^{*}$ and $F_{*}\sim
p_{d}a_{0}^{*}$ we evaluate $W' \sim
(p_{d}a_{0}^{*})^{2}/Y^{*}$. Substituting the expression of
$a_{0}^{*}$ into the expression for $W$ and $W'$ we obtain
the COR as 

\begin{equation}\label{e-plastic}
e^{2}=\frac{v_{r}^{2}}{v_{i}^{2}}=\frac{W'}{W} \sim \frac{p_{d}^{4/3}R^{2/3}}{Y^{*}(MV_{i}^2)^{1/3}} ,
\end{equation}
where $v_r$ is the rebound velocity.
Thus, we expect the law $e \sim v_{i}^{-1/3}$ in the
collision of a plastic deformed disk. The three dimensional
version of evaluation which gives $e \sim v_{i}^{-1/4}$
agrees well with the experiment\cite{johnson}.

 Our finding is, however, something new, because (i) the drop of COR
 is excited by the temperature and (ii) COR decreases more
 rapidly like $e\sim v_i^{-1.2}$ than that for the conventional plastic
 deformation $e\sim v_i^{-1/3}$ in (\ref{e-plastic}).
The mechanism how to occur the plastic deformation is not clear at
 present including the linear law in Fig.\ref{vcr}.

\section{Conclusion}

We have numerically studied the impact of a two dimensional
elastic disk with
the wall with the aid of model A and model B. 
The result can be summarized as
(i) The coefficient of restitution (COR) decreases with the impact
velocity.
(ii) The result of our simulation is not consistent with the
result of the two-dimensional
quasi-static theory. For large impact velocity, there is hysteresis in
the deformation of the center of mass. For small velocity, there remains the 
inelastic force even at $\dot h=0$.
(iii) There are drastic effects of temperature in both small and large
impact velocity.
(iv) In particular, for large impact velocity of model A,
 we have found the abrupt
drop of COR above the critical impact velocity by the
plastic deformation. The critical velocity of the plastic
deformation seems to obey a
simple linear function of temperature.

We believe that this preliminary 
report is meaningful to recognize
that physicists have poor understanding of  such the fundamental
process of elementary mechanics.
 We hope that this letter
will invite a lot
of interest in the impact from various view points.
We, at least, have a plan to  study three dimensional impacts to
clarify the relation among
the microscopic simulation, experiments and the quasi-static elastic theory. 

\vskip 0.4cm
\noindent{\bf Acknowledgment}

We appreciate S. Sasa, S. Takesue, Y.Oono and H. Tasaki
 for their valuable comments. One of 
the authors(HK) thanks S. Wada, K. Ichiki, A. Awazu, and
M. Isobe for stimulative discussions. 
This study is partially supported by  
the Grant-in-Aid for Science 
Research Fund from the Ministry of Education, Science and Culture 
(Grant No. 11740228).


\begin{thebibliography}{99}

\bibitem{goldsmith} W. Goldsmith, Impact: {\it The Theory and Physical
 Behavior of Colliding Solids} (Edward Arnold Publ., London, 1960).
\bibitem{high} At least, the subject of inelastic collisions is always 
  included in textbooks of physics in Japanese high schools.
\bibitem{granules} see .e.g. L. P. Kadanoff, Rev. Mod. Phys. {\bf 71},
 435 (1999); P. G. de Gennes, {\it ibid}, S367 
 (1999) and references therein.
\bibitem{love} A. E. H. Love, A Treatise on the Mathematical Theory of
 Elasticity (Cambridge Univ. Press, 1927).
        \bibitem{landau}{L. D. Landau and E. M. Lifshitz,
 \textit{Theory of Elasticity (2nd English ed.)}
 (Pergamon,New York, 1960)}.
        \bibitem{johnson}{K. L. Johnson,
 \textit{Contact Mechanics}
 (Cambridge University Press, Cambridge, 1985)}.
\bibitem{hills} D. A. Hills, D. Nowell and A. Sackfield, Mechanics of
 Elastic Contacts (Butterworth-Heinemann, Oxford, 1993).

\bibitem{hertz} H. Hertz, J. Reine Angew. Math. {\bf 92}, 156 (1882).
\bibitem{mindlin} R. D. Mindlin, J. Appl. Mech. Trans. ASME {\bf 16},
 259 (1949). See also ref.7.
\bibitem{dem} P. A. Cundall and O. D. L. Strack, Geotechinique, {\bf
 29}, 47 (1979).


\bibitem{newton} I. Newton, Philoshophiae naturalis Principia
 mathematica (W. Dawason and Sons, London, 1962). The original one has
 been published in 1687.
        \bibitem{sonder}{See, for example, R. Sondergaard, K. Chaney, 
and C. E. Brennen, Transaction of the ASME, 
Journal of Applied Mechanics \textbf{57}, 694 (1990);
 F. G. Bridges, A. Hatzes, and D.N.C. Lin, Nature
 \textbf{309}, 333 (1984);
 K. D. Supulver, F. G. Bridges, and D. N. C. Lin, ICARUS
 \textbf{113}, 188 (1995)} 
\bibitem{1d} However, the coefficient of restitution of the
collisions between one-dimensional rods does not depend on the colliding
velocity but 
is  determined by the ratio of
 length of the colliding rods. See ref.\cite{goldsmith}. Recent studies
 on one-dimensional collisions can be seen in
G. Giese and A. Zippelius, Phys. Rev. E {\bf 54}, 4828 (1996);
T. Aspelmeier, G. Giese and A. Zippelius, Phys. Rev. E {\bf
  57}, 857 (1998); A. G. Basile and R. S. Dumont,
Phys. Rev. E {\bf 61}, 2015 (2000).
\bibitem{kuwabara} G. Kuwabara and K. Kono, Jpn. J. Appl. Phys. {\bf
 26}, 1230 (1987).


\bibitem{brilliantov96} N. Brilliantov, F. Sphan, J.-M. Hertzsch  and
T. P\"oschel, Phys. Rev. E {\bf 53}, 5382 (1996).
\bibitem{morgado} W. A. Morgado and I. Oppenheim, Phys. Rev. E {\bf 55}, 
 1940 (1997).
\bibitem{brilliantov} e.g. T. Schwager and T. P\"oschel, Phys. Rev. E
 {\bf 57}, 650 (1998);  R. Ram\'irez, T. P\"oschel, N. Brilliantov and
 T. Schwager, Phys. Rev. E {\bf 60}, 4465 (1999).
        \bibitem{gerl}{F. Gerl and A. Zippelius,
 Phys. Rev. E \textbf{59}, 2361 (1999)}.
        \bibitem{hoover}{W. G. Hoover,
 \textit{Computational Statistical Mechanics}
 (Elsevier Science Publishers B. V., Amsterdam, 1991)}.
\end{thebibliography}
\end{document}